\documentclass[a4paper,twoside]{article}
\newcommand{\affil}[1]{$^{\rm #1}$}
\usepackage[authoryear]{natbib}
\bibpunct{(}{)}{;}{a}{}{,}
\usepackage{graphicx}
\usepackage{amsmath}
\date{}
\newcommand{\kms}{\mbox{km\,s$^{-1}$}}

\def\la{\mathrel{\mathchoice {\vcenter{\offinterlineskip\halign{\hfil
$\displaystyle##$\hfil\cr<\cr\sim\cr}}}
{\vcenter{\offinterlineskip\halign{\hfil$\textstyle##$\hfil\cr
<\cr\sim\cr}}}
{\vcenter{\offinterlineskip\halign{\hfil$\scriptstyle##$\hfil\cr
<\cr\sim\cr}}}
{\vcenter{\offinterlineskip\halign{\hfil$\scriptscriptstyle##$\hfil\cr
<\cr\sim\cr}}}}}
\def\ga{\mathrel{\mathchoice {\vcenter{\offinterlineskip\halign{\hfil
$\displaystyle##$\hfil\cr>\cr\sim\cr}}}
{\vcenter{\offinterlineskip\halign{\hfil$\textstyle##$\hfil\cr
>\cr\sim\cr}}}
{\vcenter{\offinterlineskip\halign{\hfil$\scriptstyle##$\hfil\cr
>\cr\sim\cr}}}
{\vcenter{\offinterlineskip\halign{\hfil$\scriptscriptstyle##$\hfil\cr
>\cr\sim\cr}}}}}

\DeclareMathAlphabet{\mathsc}{OT1}{cmr}{m}{sc}
\def\testbx{bx}%
\DeclareRobustCommand{\ion}[2]{%
\relax\ifmmode
\ifx\testbx\f@series
{\mathbf{#1\,\mathsc{#2}}}\else
{\mathrm{#1\,\mathsc{#2}}}\fi
\else\textup{#1\,{\mdseries\textsc{#2}}}%
\fi}

\def\aj{AJ}
\def\araa{ARA\&A}
\def\apj{ApJ}
\def\apjl{ApJ}
\def\apjs{ApJS}

\def\aap{A\&A}

\def\aaps{A\&AS}
\def\mnras{MNRAS}

\def\pasa{PASA}

\title{\large\bf\flushleft Measuring The Mass-Loss Evolution at The Tip of\\ The Asymptotic Giant Branch}
\author{\parbox{\textwidth}{\flushleft
\vspace{-0.5cm}
{\it C.\ Sandin\affil{A,B}, M.\ M.\ Roth\affil{A}, D.\ Sch{\"o}nberner\affil{A}}\\
\vspace{0.4cm}
{\small \affil{A}\,Astrophysikalisches Institut Potsdam, An der Sternwarte 16, D-14482 Potsdam, GERMANY}\\
{\small \affil{B}\,Email: CSandin@aip.de}}}
\begin{document}
\twocolumn[
\begin{changemargin}{.8cm}{.5cm}
\begin{minipage}{.9\textwidth}
\vspace{-1cm}
\maketitle
\small{\bf Abstract: }

In the final stages of stellar evolution low- to intermediate-mass stars lose their envelope in increasingly massive stellar winds. Such winds affect the interstellar medium and the galactic chemical evolution as well as the circumstellar envelope where planetary nebulae form subsequently. Characteristics of this mass loss depend on both stellar properties and properties of gas and dust in the wind formation region. In this paper we present an approach towards studies of mass loss using both observations and models, focusing on the stage where the stellar envelope is nearly empty of mass. In a recent study we measure the mass-loss evolution, and other properties, of four planetary nebulae in the Galactic Disk. Specifically we use the method of integral field spectroscopy on faint halos, which are found outside the much brighter central parts of a planetary nebula. We begin with a brief comparison between our and other observational methods to determine mass-loss rates in order to illustrate how they differ and complement each other. An advantage of our method is that it measures the gas component directly requiring no assumptions of properties of dust in the wind. Thereafter we present our observational approach in more detail in terms of its validity and its assumptions. In the second part of this paper we discuss capabilities and assumptions of current models of stellar winds. We propose and discuss improvements to such models that will allow meaningful comparisons with our observations. Currently the physically most complete models include too little mass in the model domain to permit a formation of winds with as high mass-loss rates as our observations show.


\medskip
\medskip
\end{minipage}
\end{changemargin}
]
\small

\section{Introduction}\label{sec:introduction}
Low- to intermediate-mass stars, with a main sequence mass of $0.8\,M_{\odot}\!\la\!M\!\la\!8\,M_{\odot}$, lose most of their envelopes as they ascend the thermally pulsing asymptotic giant branch (TP-AGB). The increasing rate of mass loss dominates time scales of stellar evolution as the stellar wind changes the star more rapidly than nuclear burning in the stellar core does. The matter of the stellar wind is enriched with nuclearly synthesized elements, which are dredged up from the stellar interior during the evolution. These winds are also sites of strong dust formation and determine properties of the physical environment where planetary nebulae (PNe) form subsequently. Because most stars pass through this phase mass loss from AGB stars contribute significantly to the chemical evolution of the interstellar medium and the galaxy. Recent reviews of mass loss from cool stars are given by, e.g., \citet{Wi:00}, \citet{Ho:05}, \citet{Zi:06}, and \citet{vLo:08}. Also see Zijlstra (these proceedings). The review of \citet{Ho:09} focuses on the role of dust.

Stellar winds of AGB stars form in a complex environment of many variables, here is a brief and much simplified description of the formation process. These stars tend to pulsate as long period variables, with time scales of several hundred days. The pulsations form in the outer part of a convective envelope, that surrounds a bright giant star where hydrogen and helium burn alternately in shells on longer time scales, of about $10^3$--$10^4$\,yr. The stellar atmosphere is cool and allows formation of large amounts of sub-micron- (to \mbox{micron-)} sized dust, which is either carbon-rich (C-stars) or based on other elements (M-stars and S-stars). The dust forms under non-equilibrium conditions at similar time scales as the pulsations, and dust in many circumstances provides the main driving force for the wind due to its large opacity; the dust-driven wind forms when individual dust particles drag the gas along by collisions. The wind shapes an extended stellar atmosphere, and further out a circumstellar envelope where the radial ($r$) density gradient would be proportional to $r^{-2}$ at a constant mass-loss history.

When the central star is not concealed by a massive envelope anymore it heats up and can, under the right circumstances, ionize the matter of the AGB wind. The shape of the then formed ionized (planetary) nebula shows a variety of geometries \citep[see e.g.][]{BaFr:02,SaMoSa.:07}. In view of PNe with a spherical or elliptical geometry they consist of a central part with a relatively bright rim and a fainter attached shell; both are built-up in an interplay between photo-ionization and wind interaction \citep[see, for example,][]{FrBaRi:90,MaSc:91,Fr:94,Me:94,PeScSt.:04}. The matter in the central parts of these objects has been restructured during their formation and this mass can therefore not be used to measure properties of the last period of more spherically symmetric mass loss on the AGB. As the PN evolves the region outside the PN proper is eventually ionized, and a halo forms (also see Sect.~\ref{sec:halos}). Such halos are relatively unaffected by processes of the formation of the PN, and therefore still contain directly accessible information of the previous AGB mass loss. \citet{So:06b} discusses the transition from more spherically symmetric outflow on the AGB to axi-symmetric outflow that is often observed in later stages.

Measurements of mass loss from AGB stars are made by four different methods, using spectroscopy, imaging, or both. Observations, which cover and spatially resolve the circumstellar envelope, readily provide information on the mass-loss evolution throughout the envelope. 

In one spectroscopic method radio emission lines of carbon monoxide are compared with models of radiative transfer in the circumstellar envelope. Mass-loss rates are hereby calculated using best-estimate fits; \citet{Sc:07} reviews this method. The reliability of the outcome of this method, that measures the gas-component in the wind, is discussed by e.g.\ \citet{Sa:90}, \citet{Ka:92}, and \citet{ScOl:01}. This approach works best with mass-loss rates $\dot{M}\!\la\!10^{-5}\,\text{M}_{\odot}\text{yr}^{-1}$ \citep[at higher values the relevant lines become saturated;][]{RaScOl.:08}.

In another spectroscopic method infrared measurements of dust are used to match the observed spectral energy distribution with models of distribution of dust; this method is reviewed by \citet{vLo:07}. Different models are used to prescripe input mass-loss rates, ranging from constant values to mass-loss histories \citep[see e.g.][]{Gr:95b}. In order to fit the spectral energy distribution of some stars in the Magellanic Clouds it seems necessary to vary the model input mass-loss rate by a factor of a few over time periods of $10^2$--$10^3$\,yr \citep[although as the authors comment this should be a less likely scenario]{GrWoSl.:07}. Both spectroscopic methods require additional assumptions of properties of the mass-loss rate, viz.\ the expansion velocity, and the dust-to-gas mass ratio.

\citet{MMa:09} reviews imaging methods for measuring the mass loss, the chemistry, etc., using filters covering ultra-violet to infrared wavelengths.

In a fourth method \citet{SaScRo.:08} study four halos of Galactic Disk PNe and measure the mass-loss evolution and other properties after the last thermal pulse near the tip of the AGB. Our new approach uses the novel method of integral field spectroscopy (IFS), that in principle is a low spatial resolution imaging method. Based on the approach of \citet{PlSo:90} this method allows the radial gas density structure -- i.e.\ the mass-loss evolution -- to be measured directly. No additional fitting procedure is necessary using supplementary models, as in the other two spectroscopic methods. Lacking better data, however, a constant value is currently used with the expansion velocity. This method, furthermore, only works when observing the final stages of AGB mass loss since the matter only becomes ionized during the PN phase. It obviously also only works with stars which form a PN (and later also a halo). The authors find that the mass-loss evolution shows increased values by a factor of about 4--7 during $10^4\,$yr for all four objects.

\citet[also see \citealt{Pe:81}]{Re:81} introduces the term \emph{superwind} in order to emphasize that stars ascending the AGB reach higher mass-loss rates than stars on the precedent red giant branch. Using stellar-evolution calculations including an empirical mass-loss relation (that depends on the pulsation period, but not the stellar mass or the luminosity) \citet{VaWo:93} find that mass-loss rates reach the highest values during the final thermal pulses on the AGB. \citet[who also provides a summary of previously derived mass-loss relations]{Bl:95} uses a mass-loss relation for the AGB, that is based on the red-giant branch relation of \citet[although accounting for effects of the dynamical models of \citealt{Bo:88}]{Re:75}, and thereby also accounts for increasing mass-loss rates in his stellar-evolution calculations. \citet{ScWiSe:99} reach the same conclusion, using a mass-loss relation that is based on models of dust-driven winds \citep[adopting the relation of][]{ArFlSe:97}. \citet{vLoGrKo.:99,vLoCiZi.:05}, furthermore, come to a similar conclusion with their observational studies of dust-enshrouded stars. \citet{HeFoOm.:90} and \citet{JuSkTi.:96} present additional results which support the above findings. Using time-dependendent photoionization models of PNe \citet{ScJaSt.:05}, moreover, find that the AGB wind must increase in strength during the final evolution. Models assuming a stationary AGB wind fail to explain virtually all observed PNe.

Due to the small number of assumptions our observational method provides a powerful approach to study the decisive stage of stellar evolution with strong mass loss at the tip of the AGB. In this paper we present the key concepts of our method, including suggestions on how to use it with additional objects. We also discuss required improvements to models of stellar winds to be used when interpreting the outcome of our observations. We first describe properties and observables of the objects of our observational method, AGB halos of PNe, in Section~\ref{sec:halos}. Thereafter we discuss assumptions and applicability of current models of stellar winds in Section~\ref{sec:modeling}.

\section{About AGB halos of PNe}\label{sec:halos}
The definition of what a PN halo is comes from photo-ionization models of PNe, which have a spherical or elliptical shape. The halo is identified as the ionized region outside the central PN \citep[see e.g.][]{StSc:03,ViGaMa:02,ViMaGa:02}. It appears a few thousand years after the central PN becomes visible. The halo is also less affected by PN shaping processes than the central region is, and therefore allows direct measurements of properties of the AGB wind. However, in general the least affected region of mass loss in the halo is the inner part. In the outer halo it is expected that wind interaction, between periods of (the last) thermal pulse(s) and intermediate periods of mass loss, has modified the flow. The interaction between the PN and the interstellar medium (ISM) is studied by e.g.\ \citet{ViGaMa:03} and \citet{WaZiOb:07}. They find that the nebula can become asymmetric when the spherically symmetric structure of an AGB wind moves through the ISM. For such objects \citep[which fraction is about 20\% of all objects,][]{WaZiOb:07} it is important to measure the mass loss in the wake of the interaction.

\subsection{How to differ between AGB halos and other structures}\label{sec:AGBhalo}
The term halo has in the past often been used to indicate different types of structures. \citet[cf.\ Section 3.2 therein]{CoScSt.:03} require four criteria to be met in order to positively identify the presence of an AGB halo. We use the same criteria, but rephrase them slightly as follows:
\begin{itemize}
\item The halo is limb-brightened, or \emph{detached} according to the definition of \citet{ChJaAr:87}.
\item The surface brightness (SB) of the halo must be of the order $10^2$--$10^4$ weaker than the peak emission on the rim in the inner nebula, or even weaker. The intensity of the halo decreases with increasing radius, and halos of some objects show this full intensity range.
\item The radius of the halo is at least twice as large as the inner rim radius.
\item Objects which appear to be evolved and have a low-luminosity central star together with an average halo-SB of about $10^{-2}$ of the peak intensity are candidates to have a recombination halo \citep[cf.][]{Ty:86,CoScSt.:00}. Such regions are not fully ionized anymore, and in comparison to younger PNe the matter in the halo has been restructured during the evolution of the PN. These objects provide little information about the AGB wind.
\end{itemize}
The bright central parts of a PN may influence measurements of the halo around it through instrumental scattered light. \citet{MiClWa:89b}, \citet{CoScSt.:03}, and \citet{SaScRo.:08} present methods to distinguish halos from instrumental scattered light.

\citet{CoScSt.:03} find 25 Galactic Disk AGB halos, which are either circular, slightly elliptical, or highly asymmetric. This is about 60\% of the sample of elliptical objects in their study. \citet{SaScRo.:08} find two additional halos; \citet{CoScSt.:03} mark one of these objects, IC\,3568 to not have a halo. \citet[SMP27]{RePa:06} and Cohen (these proceedings) identify two additional AGB halos in the Magellanic clouds. Although relatively few AGB halos are known it appears that they are a common phenomenon, where their detection is mainly limited by their brightness.

\subsection{About rings found around AGB stars, post-AGB stars and PNe}\label{sec:rings}
In many cases images of AGB stars, post-AGB stars and PNe reveal multiple ring-shaped shells (also called arcs) around the main object. These shells are separated by time periods of about $10^2$--$10^3\,$yr, see e.g.\ \citet{SaTrWa.:98,SaZiBu.:99}, \citet{KwSuHr:98}, \citet{SuVoKw.:98}, \citet{MaHu:99,MaHu:00,MaHu:06}, \citet{Bo:00}, \citet{HrKwSu:01}, \citet{TeHa:02}, and \citet{CoScSt.:03,CoSaMe.:04}. The origin of the rings is not understood, but numerous processes have been suggested to explain their formation. For example, \citet{MaMo:99} argue that the rings result from binary interaction, \citet{SiIcDo:01} advocate an origin from a drift-induced instability, \citet{ZiBe:02} suggest a pulsational instability as the origin, and \citet{vHoThFr.:03} argue for a nuclear-burning instability. A more complete discussion of these and additional suggested mechanisms is given by \citet{So:02,So:06b}. Our method to measure mass-loss rates using IFS results in data with too low spatial resolution, in general, to find these rings \citep[see, however,][for NGC\,3242]{MoRoSc.:05}. Although the rings arise due to stellar processes, which are important to understand, their influence on measurements of the mass-loss evolution currently appears minute. In order to measure and evaluate density differences of matter, which is present in the rings and between the rings, it is necessary to further increase the spatial resolution of the mass-loss measurements.

\subsection{Observables in the halo for the optical wavelength range}\label{sec:observables}
The low surface brightness of AGB halos implies a lower number of observable emission lines than in the bright central PN. Visible emission lines in the optical wavelength range, which span most of the radial extent of the halo, include the Balmer lines H$\alpha$ and H$\beta$, the two bright oxygen lines $[\ion{O}{iii}]\,\lambda\lambda4959,\,5007$, and often also neon $[\ion{Ne}{iii}]\,\lambda3869,\,3968$ ($[\ion{Ne}{iii}]\,\lambda3968$ may be blended by H$\epsilon$). In the innermost part of the halo the number of visible lines is larger. Depending on the density and the ionization structure some of the density-sensitive lines $[\ion{O}{ii}]\,\lambda\lambda3726,\,3729$, $[\ion{Ar}{iv}]\,\lambda\lambda4711,\,4740$, and $[\ion{S}{ii}]\,\lambda\lambda6717,\,6731$, are normally visible. The weak temperature-sensitive aureole line $[\ion{O}{iii}]\,\lambda4363$ is mostly visible, although depending on the site it may be blended by a much stronger telluric mercury line Hg$\,\lambda4358$. Additional emission lines, that are often visible, include H$\gamma$, H$\delta$, $\ion{He}{i}\,\lambda4471$, $\ion{He}{ii}\,\lambda4686$, and possibly $[\ion{N}{ii}]\,\lambda\lambda6548,\,6583$.

The low density throughout the halo requires it to be measured using recombination lines. Hereby the strongest Balmer lines can be used to calculate the electron density in the halo (assuming a distance to the object) and thereafter the gas density and the mass-loss rate. The line ratios of $[\ion{O}{ii}]$, $[\ion{Ar}{iv}]$, and $[\ion{S}{ii}]$ can be used to scale the density structure using values measured near the inner boundary of the halo. In our study it turned out that the recombination-line density structure matches the line-ratio densities only for one object assuming a proper distance (viz.\ NGC\,6826). For the remaining three objects it is, for some reason, necessary to use much shorter distances when calculating a recombination-line density structure that matches the line-ratio density measurements. It therefore appears, that because of this scaling it is, in general, not possible to determine an object distance.

The oxygen lines are normally used to calculate the electron temperature in both the central PN and in the halo; the temperature-sensitive nitrogen line is too weak in the halo. Detailed radial structures of the temperature in several cases reveal hot halos \citep[such tendencies are also identified by \citealt{MiClWa:89,MiClWa.:91,MaPo:89,MoRoSc.:05,KrCo:05}]{SaScRo.:08}. \citet{Ma:93} finds a simple explanation to the high temperatures with non-equilibrium models which account for hydrodynamic effects. With such models the ionization front, that is responsible for the high temperature in the halo, moves through the envelope in a short time period (typically $10^2$--$10^3$ years long). These models suggest that high temperatures in the halo are a fairly common, although transient phenomenon \citep[cf.\ Sect.\ 6.1 in][]{SaScRo.:08}. Accurate electron temperatures are necessary in order to measure abundances in the halo correctly -- although, unless additional emission lines at either shorter or longer wavelength ranges are used it is difficult to get such values for many elements.

The four mass-loss evolution structures that \citet{SaScRo.:08} measure assume a constant outflow velocity of $10\,\kms$. Mass-loss estimates will be more accurate if a radial outflow velocity structure could be measured also in the halo. The resulting line broadening, due to the expected velocities, is moderate. Nevertheless the approach of \citet{GuViMa:98} could be useful to make such measurements. Note that radial density and temperature structures, that are both provided by our method, are required in order to deconvolve kinematic measurements when determining a radial velocity structure.

\section{Modeling AGB mass loss}\label{sec:modeling}
A detailed theory of formation of stellar winds from cool stars should be able to reproduce observational findings. Because there are so many relevant physical processes in the AGB wind -- where many processes are not well understood -- a complete and self-consistent modeling of the AGB wind is (still) difficult to make.

A complete model of mass loss on the AGB should be three-dimensional. It should cover a radial region extending from the core (and possibly even include the core) out into, and including, the wind formation region, say, out to about 50 stellar radii. It should, moreover, properly handle the convective mantle between the core and the atmosphere, and thereby be able to describe the formation of stellar pulsations. It should handle dust formation in different chemistries as well as the gas-dust interaction. Radiative transfer should be done in a frequency-dependent approach accounting for possible non-local thermal equilibrium. And the model should work for a significant period of the evolution of the AGB star.

\subsection{State of current wind models}\label{sec:currentmodels}
The formidable problem of stellar wind formation can only be solved by breaking it down into manageable pieces, often with large -- and necessary -- simplifications. Here follows a general list of such key pieces, which have contributed to the development of the physical part of models of mass loss so far. The first time-dependent stellar wind models that can account for effects of pulsations are presented by \citet{Wo:79} and \citet[who also, because of a lack of the necessary data at the time, introduces a constant gas opacity, that still is in use]{Bo:88}. \citet{FlGaSe:92} and \citet{HoFeDo:95} present time-dependent wind models, which use the time-dependent dust formation description of \citet[for C-rich dust; also see the references therein]{GaGaSe:90}. \citet{HoJoLoAr:98} and \citet{HeWiSe:00} introduce gray molecular opacities, which replace the constant opacity. \citet{GaSe:99} extend the dust formation description to work in M-type stars. \citet{SaHo:03} introduce the physics and role of gas-dust drift -- although the physical outcome is somewhat affected by numerical issues (which are now corrected, see below). \citet{HoGaArJo:03} introduce frequency-dependent opacities for both the gas and the dust. \citet{JeWi.:03} study dust formation in M-type stars. \citet{SaHo:04} find that gas-dust drift increases the amount of formed dust. Moreover, \citet{WoNi:05} and \citet{Wo:06} study the outcome of two-dimensional wind models. \citet{Sa:08} improves the numerical accuracy of both drift and non-drift models (cf.\ Section~\ref{sec:nummodeling}). \citet{FrHo:08} present a first step towards three-dimensional wind models, and \citet{Ho:08} finds that a frequency-dependent approach and larger (micron-sized) grains may provide the necessary force to drive a wind off M-type stars.

Self-consistent models of AGB mass loss should rely on a small number of input parameters -- the mass-loss rate and the expansion velocity must not be such parameters. Current models, as described by e.g.\ \citet{Sa:08}, require the following four parameters to be specified: the stellar luminosity, the effective temperature, the composition (in this case the C/O-ratio), and the stellar mass. In one-dimensional models it is also necessary to describe the stellar pulsations using either a description for turbulent energy (i.e.\ convection) or a moving boundary (a so-called piston). This adds about 2 parameters (cf.\ Section~\ref{sec:finalmodeling}).

There are, however, more assumptions built into the models. The gas is described by the equation of state, assuming an ideal gas with a mean molecular weight. The interaction between the gas and the radiation field is described by opacities, which are calculated in advance for the necessary atoms and molecules. Although the gas opacity does not provide a strong enough force to drive a wind it is a decisive component that determines the density gradient in the atmosphere -- and thereby influences the wind formation indirectly. In a sense, using a constant gas opacity all other detailed physics becomes of secondary importance. A number of simplified mass-loss rate formulae are in use, which are based on mass-loss models using a constant gas opacity; see e.g.\ \citet[whose mass-loss relation is used by \citealt{ScWiSe:99} in their stellar-evolution calculations]{ArFlSe:97}, \citet{WaScWi.:02}, and \citet{WaWiSc.:08}. Because models using a constant gas opacity achieve unrealistic density structures, these formulae cannot be expected to provide quantitatively useful estimates of mass-loss rates. Based on this argument alone the mass-loss relation of \citet{MaWaHo:09}, that is provided in the form of an interpolating computer program, seems more appropriate as it is inferred using frequency-dependent opacities.

The dust component is, furthermore, likewise described by properties of the dust grains and through the dust opacity; in their parameter study \citet{AnHoGa:03} find that wind models are highly sensitive to how such dust properties are used. The term describing the interaction between gas and dust, the drag force, is rather insensitive to which type of collisions are used (specular or diffusive). \citet{SiIcDo:01} find that a gas-dust drift-instability is the cause of the rings observed around many (former) AGB stars (cf.\ Section~\ref{sec:rings}). Their models are, however, very difficult to reproduce as they are based on a description using several simplifications \citep[this is also noted by][]{So:02}. The inner boundary is hydrostatic, thus pulsations are not considered. They treat radiative transfer in a simplified approach assuming an optically thin envelope, and use only one parameter setup of rather unrealistic properties. An extended period of mass loss is modeled by replenishing mass at the inner boundary (by keeping the gas density fixed). Instead of reproducing this setup we propose to make the required physical enhancements to our models (see below) and thereafter see if and how a similar drift-instability appears.

\subsection{Effects on models of improved numerical accuracy}\label{sec:nummodeling}
The numerical description in AGB star wind models is explored in detail by \citet{Sa:08}. He introduces and studies several improvements to the numerical approach. The results show that most one-dimensional wind structures are steady, with small periodic fluctuations. In comparison a majority of previous wind model structures are mildly chaotic. The formation of dusty shells in the wind, that were the topic of many papers up to now, do not form in the new models. Although, models which still use a constant gas opacity instead of molecular opacities continue to form chaotic structures; probably as a consequence of the fewer degrees of freedom which are available in such models.

Models which are calculated using gas-dust drift require a more careful numerical treatment than non-drift models do. It turns out that the dust is more tightly coupled to the radiation field than it is to the gas component. The dust thereby pulsates markedly along with the modulated radiation field. It should be remembered that in comparison to real (three-dimensi\-onal) stars these models are one-dimensional. Effects, such as convection and formation of clumps, which could cause an inhomogeneous or even chaotic flow, are not considered in these models.

\subsection{Modeling mass-loss formation\\ near the tip of the AGB}\label{sec:finalmodeling}
In order to model stellar winds near the tip of the AGB it is necessary to enhance current wind models. Normally a radial domain is used that extends from above the region where pulsations form (at about 0.8--0.9 stellar radii) to 10--60 stellar radii. The amount of mass present in the model within this domain (about $10^{-3}$--$10^{-2}\,\text{M}_{\odot}$, using gray or frequency-dependent opacities, respectively\footnote{In comparison most of the stellar mass is placed in the envelope in models which use a constant gas opacity, and the standard value $\kappa_\text{g}\!=\!2\!\times\!10^{-4}\,\text{cm}^2\text{g}^{-1}$.}) is not enough to calculate mass-loss rates of the order $\ga\!10^{-5}\,\text{M}_{\odot}\text{yr}^{-1}$ with any reliability. For example, it normally takes about $10^2$ years of evolution to start a dynamic model from a hydrostatic initial model.

By including more of the central parts of the star in the model domain it becomes possible to model also the highest mass-loss rates -- and thereby provide a mean to understand our observations at the tip of the AGB. The inclusion of the convective stellar mantle in the model requires the use of either a three-dimensional approach \citep[such as presented by][]{FrHo:08} or a description for turbulent energy \citep[see e.g.][]{Ku:86,GeWi:92,CaDu:98,WuFe:98,Fe:99a}, which implies a number of additional free parameters. It should, however, be possible to set many of these new parameters to default values. \citet{So:06b} also suggests an approach where more of the entire star is modeled in order to fully understand the formation and role of the pulsations. Finally, it may also turn out that dust cannot itself drive the strongest winds observed, and in this case it may be necessary to find another, complementary, driving mechanism.

\section*{Acknowledgments}
C.~S.\ is partly supported by DFG under grant number SCHO 394/26. We thank M.~Steffen for a valuable discussion and an anonymous referee for several constructive comments, which have helped us to improve our presentation.

\end{document}